\documentclass[aps,prd,10pt,twocolumn,preprintnumbers,amsmath,amssymb,nofootinbib,floatfix]{revtex4-2}
\usepackage{graphicx,amsfonts,booktabs,multirow,slashed,braket,bm,xcolor}
\usepackage{lmodern}
\usepackage[colorlinks=true,citecolor=blue,linkcolor=red,urlcolor=blue]{hyperref}
\usepackage[capitalize,nameinlink]{cleveref}
\usepackage[expansion=false]{microtype}
\usepackage{orcidlink}

\newcommand{\dd}{\mathrm d}
\newcommand{\hc}{\mathrm{H.c.}}
\newcommand{\tju}{\affiliation{Center for Joint Quantum Studies and Department of Physics, School of Science, Tianjin University, Tianjin 300350, China}}

\allowdisplaybreaks[4]

\begin{document}
\title{Production of hidden-charm molecular candidates with hidden strangeness in \texorpdfstring{$B_s^0$}{Bs0} decays}
\author{Yue Su\,\orcidlink{0009-0003-0728-5483}}\tju
\author{Ying-Bo He\,\orcidlink{0009-0009-9469-7227}}\email{heyingbo@tju.edu.cn}\tju
\author{Xiao-Yun Wang\,\orcidlink{0009-0006-6091-6944}}\email{xyun@tju.edu.cn}\tju
\author{Xiao-Hai Liu\,\orcidlink{0000-0002-6159-3140}}\email{xiaohai.liu@tju.edu.cn}\tju

\date{\today}
\begin{abstract}
We study the production of spin-one $D_{s0}^*(2317)\bar D_s^*$, $D_{s1}(2460)\bar D_s^*$, and $D_{s1}(2536)\bar D_s^*$ molecular candidates in $B_s^0$ decays with an accompanying $\eta$, providing a complementary probe of their structure and production mechanisms. The meson-loop rescattering amplitudes are constructed from factorized weak sources and heavy-hadron chiral interactions. Constituent couplings are estimated using threshold-dependent compositeness relations, and the associated input uncertainties are propagated to the predicted rates. Opposite-$C$-parity partner hypotheses are included with inherited benchmark parameters. We calculate the branching fractions for two relative signs. Over the cutoff range $2\leq\alpha\leq5$, the larger central branching fractions reach several $10^{-6}$ to several $10^{-5}$, suggesting potential accessibility in future experimental measurements.
\end{abstract}
\maketitle
\raggedbottom

\section{Introduction}
The growing spectrum of charmoniumlike states has challenged a description based only on conventional $c\bar c$ mesons. Proposed explanations include compact multiquark states, hadronic molecules, hybrid mesons, and threshold effects; their experimental signatures and theoretical interpretations are reviewed in Refs.~\cite{Chen:2016qju,Brambilla:2019esw,Chen:2022asf,Guo:2017jvc,Liu:2019zoy,Meng:2022ozq,Bai:2026atm,Wang:2025dur,Wang:2025sic}.

Representative examples are the $X(3872)$ observed in $B$ decays~\cite{Belle:2003nnu}, the charged $Z_c(3900)$ seen in the $\pi J/\psi$ channel~\cite{BESIII:2013ris,Belle:2013yex}, and the vector structure historically denoted $Y(4260)$ in $\pi^+\pi^-J/\psi$~\cite{BaBar:2005hhc,CLEO:2006tct}. Angular analyses established $J^{PC}=1^{++}$ for the $X(3872)$ and $J^P=1^+$ for the $Z_c(3900)$~\cite{LHCb:2013kgk,BESIII:2017bua}. A subsequent BESIII energy scan resolved a narrower structure near $4.22$ GeV in the $\pi^+\pi^-J/\psi$ cross section, refining the earlier $Y(4260)$ picture~\cite{BESIII:2016bnd}.

Widely studied molecular descriptions of this sector involve nonstrange open-charm mesons, such as $D\bar D^*$ and $D\bar D_1(2420)$~\cite{Guo:2017jvc,Wang:2013cya,Wang:2025clb}. Extending these studies to systems with both hidden charm and hidden strangeness tests how the light-quark flavor affects exotic spectroscopy.

The $c\bar c s\bar s$ sector can be organized either as compact tetraquarks or as pairs of charm-strange hadrons. Early evidence for a near-threshold $J/\psi\phi$ structure was reported by CDF in $B^+\to J/\psi\phi K^+$ decays~\cite{CDF:2009jgo}. The $J/\psi\phi$ structures reported in LHCb amplitude analyses of $B^+\to J/\psi\phi K^+$, including the $X(4140)$ and $X(4274)$, provide experimental motivation for this sector~\cite{LHCb:2016axx,LHCb:2016nsl,LHCb:2021uow}. These measurements illustrate the role of amplitude analyses of heavy-hadron decays in establishing the quantum numbers of exotic candidates~\cite{Johnson:2024omq}. Compact $cs\bar c\bar s$ spectra have been investigated using QCD sum rules~\cite{Chen:2016oma,Wang:2023jaw}, while meson-exchange and contact-interaction studies predict hidden-charm molecular states involving charm-strange mesons~\cite{Shen:2010ky,Dong:2021juy}. Non-resonance interpretation has also been suggested for some of these exotic candidates~\cite{Liu:2016onn}. These approaches also motivate searches for negative-parity states built from ground-state and positive-parity charmed mesons.

Several structures near $4.5$--$4.7$ GeV are directly relevant to such searches. BESIII observed the vector charmonium-like state $Y(4500)$ in the $e^+e^-\to K^+K^-J/\psi$ cross section, with a mass near $4.485$ GeV and a width of about $0.11$ GeV~\cite{BESIII:2022joj}. Its proximity to the $D_s\bar D_{s1}(2536)$ threshold has motivated a molecular interpretation connected by heavy-quark spin and $SU(3)$-flavor symmetries to the nonstrange $D\bar D_1$ sector~\cite{Peng:2022nrj}. Other explanations include a higher charmonium with $5S$--$4D$ mixing~\cite{Wang:2022jxj} and a threshold enhancement generated by triangle singularities in $Z_{cs}^{(\prime)}$-mediated rescattering~\cite{Wang:2025zbv}. Production and decay observables can help distinguish these different dynamical origins.

Belle reported the vector $Y(4626)$ in $e^+e^-\to D_s^+D_{s1}(2536)^-+\mathrm{c.c.}$ through initial-state radiation, measuring a mass near $4.626$ GeV and a width of about $50$ MeV~\cite{Belle:2019qoi}. A coupled-channel meson-exchange calculation describes this structure as a $D_s^*\bar D_{s1}(2536)$ molecule coupled to the observed $D_s\bar D_{s1}(2536)$ channel~\cite{He:2019csk}. A $P$-wave $[cs][\bar c\bar s]$ tetraquark interpretation has also been proposed in color flux-tube and QCD sum-rule calculations~\cite{Deng:2019dbg,Zhang:2020gtx}.

In the same mass region, LHCb observed the $X(4630)$ in the $J/\psi\phi$ subsystem of $B^+\to J/\psi\phi K^+$. Its analysis favors $J^P=1^-$ over $2^-$, giving the spin-exotic assignment $J^{PC}=1^{-+}$ when combined with the positive $C$ parity of $J/\psi\phi$~\cite{LHCb:2021uow}. The $X(4630)$ has been studied as a $D_s^*\bar D_{s1}(2536)$ molecule in a one-boson-exchange model~\cite{Yang:2021sue}. QCD sum-rule analyses have also considered molecular and compact $c\bar c s\bar s$ configurations with $1^{-+}$~\cite{Wang:2021ghk,Wang:2023jaw}. Thus, both vector and spin-exotic candidates motivate a common study of their charm-strange constituent channels.

The production mechanism of hadrons is closely related to their intrinsic structures. 
Weak $B_s$ decays offer a useful production mechanism for these hidden-charm and hidden-strange systems. The color-favored transition $\bar b\to\bar c c\bar s$ in $B_s^0=\bar b s$, together with the spectator $s$ quark, supplies the $c\bar c s\bar s$ flavor content. Charm-strange meson pairs produced at the weak vertex can rescatter into molecular candidates~\cite{Wu:2023fyh,Yu:2024mol}. Here we investigate the $D_{s0}^*(2317)\bar D_s^*$, $D_{s1}(2460)\bar D_s^*$, and $D_{s1}(2536)\bar D_s^*$ configurations coupled in $S$ wave to spin one. Their two charge-conjugation combinations give $J^{PC}=1^{--}$ and $1^{-+}$, providing a constituent framework for examining the $Y(4500)$, $Y(4626)$, and $X(4630)$ mass regions.

The contributions of rescattering processes in weak decays are also widely discussed in the literature~\cite{Colangelo:2002mj,Colangelo:2003sa,Cheng:2004ru,Cheng:2005bg,Lu:2005mx,Wu:2021cyc,Liu:2020orv,Han:2021azw,Ge:2022dsp,Liu:2016onn,Liu:2017vsf,Wu:2023fyh,Yu:2024mol,Li:2023hpk}.
The processes $B_s^0\to\eta X(4630)$ and $B_s^0\to\eta Y(4626)$ have been studied by Yu, Wu, and Chen in the $D_{s1}(2536)\bar D_s^*$ molecular picture~\cite{Yu:2024mol}. They obtained branching fractions of order $10^{-5}$ from meson-loop rescattering. Here we extend the constituent comparison to $D_{s0}^*(2317)\bar D_s^*$ and $D_{s1}(2460)\bar D_s^*$, resolve the two charge components of each retained topology, and examine how their coherent projections and relative source signs affect production. This comparison connects the production rates with alternative constituent assignments in the same mass regions.

The positive-parity constituents also connect this study with multihadron dynamics. The $D_{s0}^*(2317)$ and $D_{s1}(2460)$ lie below the $DK$ and $D^*K$ thresholds~\cite{BaBar:2003oey,CLEO:2003ggt}, and chiral unitary and lattice studies support substantial $D^{(*)}K$ components~\cite{Guo:2006fu,Mohler:2013rwa,Lang:2014yfa,MartinezTorres:2014kpc}. Their compositeness and its relation to scattering observables have also been examined in a coupled quark--hadron description~\cite{Terashima:2025tbz}. Correlating either constituent with $\bar D_s^*$ therefore suggests the cluster configurations $(DK)\bar D_s^*$ and $(D^*K)\bar D_s^*$. The connection between two-hadron molecular assignments and related three-body systems is reviewed in Ref.~\cite{Liu:2024uxn}. Related three- and four-hadron systems have been explored in Refs.~\cite{MartinezTorres:2009xb,Pan:2023zkl,Wu:2025fzx,Pan:2025vyg}. We describe the positive-parity constituents by effective meson fields in the production amplitudes.

In this work, we calculate $B_s^0\to\eta\mathbb{X}$ through charm-strange meson triangle rescattering diagrams, using factorized weak vertices and covariant light-front form factors~\cite{Cheng:2003sm,Yang:2024qij}. Above-threshold molecular couplings are estimated from effective-range compositeness and width saturation~\cite{Guo:2015daa,Gao:2018jhk,Guo:2019kdc,Guo:2020vmu}, while subthreshold assignments use Weinberg's weak-binding relation~\cite{Weinberg:1965hj}. Since weak $B_s^0$ decays do not conserve $C$, both molecular charge-conjugation combinations are retained. Alongside the experimentally anchored assignments, we consider opposite-$C$ partners at the same benchmark masses and widths.
We also compare individual triangle contributions and charge-projected branching fractions, including their dependence on the weak-source phases and molecular couplings. Section~\ref{sec:framework} presents the framework, Sec.~\ref{sec:results} gives the results, and the appendices collect the vertex tensors and loop contractions.


\section{Theoretical framework}\label{sec:framework}
\subsection{Molecular channels and production diagrams}
We use the following meson conventions
\begin{equation}\begin{aligned}
D_{s0}^*&\equiv D_{s0}^*(2317),\\
D_{s1}'&\equiv D_{s1}(2460),\quad D_{s1}\equiv D_{s1}(2536).
\end{aligned}\label{eq:particles}\end{equation}
The molecular fields $\mathbb{X}$, $\mathbb{X}'$ and $\widetilde{\mathbb{X}}$ denote the spin-one $D_{s0}^*\bar D_s^*$, $D_{s1}'\bar D_s^*$ and $D_{s1}\bar D_s^*$ configurations. We choose the charged-field phases
\begin{equation}\begin{aligned}
\mathsf C D_s^+&=D_s^-,&\mathsf C D_{s0}^{*+}&=D_{s0}^{*-},\\
\mathsf C D_{s1}^{(\prime)+}&=D_{s1}^{(\prime)-},&\mathsf C D_s^{*+}&=-D_s^{*-},
\end{aligned}\label{eq:Cphases}\end{equation}
so that the normalized charge states are
\begin{equation}\begin{aligned}
|\mathbb{X}^{(C)}\rangle&=\frac{|D_{s0}^{*+}D_s^{*-}\rangle-C|D_{s0}^{*-}D_s^{*+}\rangle}{\sqrt2},\\
|\mathbb{X}^{\prime(C)}\rangle&=\frac{|D_{s1}^{\prime+}D_s^{*-}\rangle-C|D_{s1}^{\prime-}D_s^{*+}\rangle}{\sqrt2},\\
|\widetilde{\mathbb{X}}^{(C)}\rangle&=\frac{|D_{s1}^{+}D_s^{*-}\rangle-C|D_{s1}^{-}D_s^{*+}\rangle}{\sqrt2}.
\end{aligned}\label{eq:Xmolwave}\end{equation}
The axial and vector constituent spins are coupled to total spin one. The $C=-1$ and $+1$ combinations have $J^{PC}=1^{--}$ and $1^{-+}$, respectively.

\begin{figure}[htbp]\centering
\includegraphics[width=.95\columnwidth]{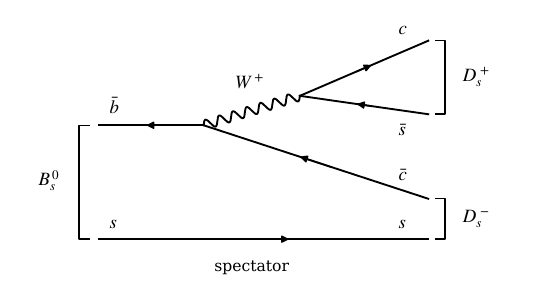}
\caption{Quark-level diagram of the Cabibbo-favored, color-allowed external-$W$-emission topology for $B_s^0 \to D_s^- D_s^+$. The transition $\bar{b} \to \bar{c} W^+$ is followed by $W^+ \to c \bar{s}$. The spectator $s$ quark combines with $\bar{c}$ to form $D_s^-$, while $c$ and $\bar{s}$ form $D_s^+$; the same quark-level pattern is used for the other charmed-strange two-meson sources in the triangle amplitudes.}
\label{fig:weakemission}
\end{figure}

The weak two-meson sources feeding the rescattering amplitudes follow the quark-level external-emission pattern shown in Fig.~\ref{fig:weakemission}. For the representative $B_s^0\to D_s^-D_s^+$ source, the spectator $s$ quark is retained in $D_s^-$, while the $c\bar s$ pair from $W^+$ decay forms $D_s^+$. The transition and emitted mesons are replaced by their vector or positive-parity counterparts in the other sources considered below.

\begin{figure}[htbp]
  \centering
  \includegraphics[width=\columnwidth]{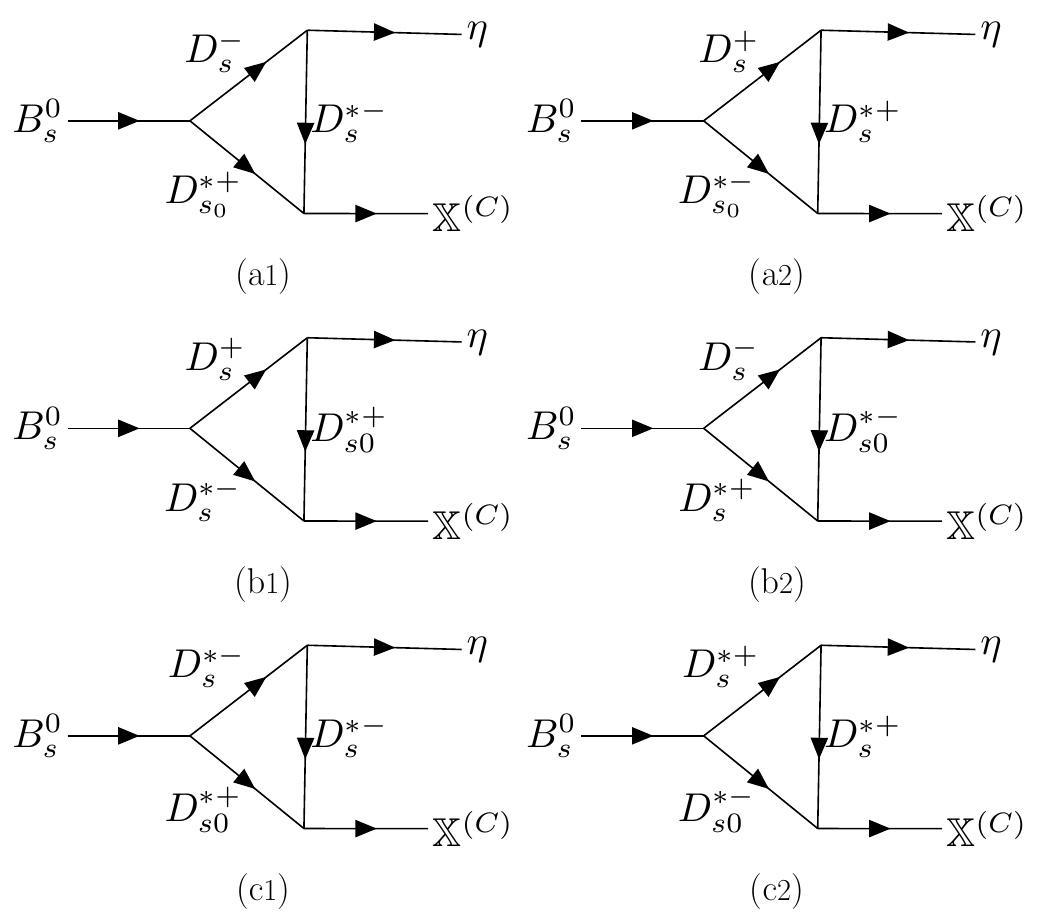}
  \caption{Triangle diagrams for $B_s^0\to\eta\mathbb{X}^{(C)}$. Panels (a1), (b1), and (c1) produce the $D_{s0}^{*+}D_s^{*-}$
    component, and panels (a2), (b2), and (c2) produce its charge
    conjugate, $D_{s0}^{*-}D_s^{*+}$.
    We denote the momenta of the upper and lower internal lines
    attached to the $B_s^0$ weak vertex by $r$ and $q$, respectively,
    and the momentum of the vertical exchanged line by $\ell$.}
  \label{fig:Ds0loops}
\end{figure}

\begin{figure}[htbp]
  \centering
  \includegraphics[width=\columnwidth]{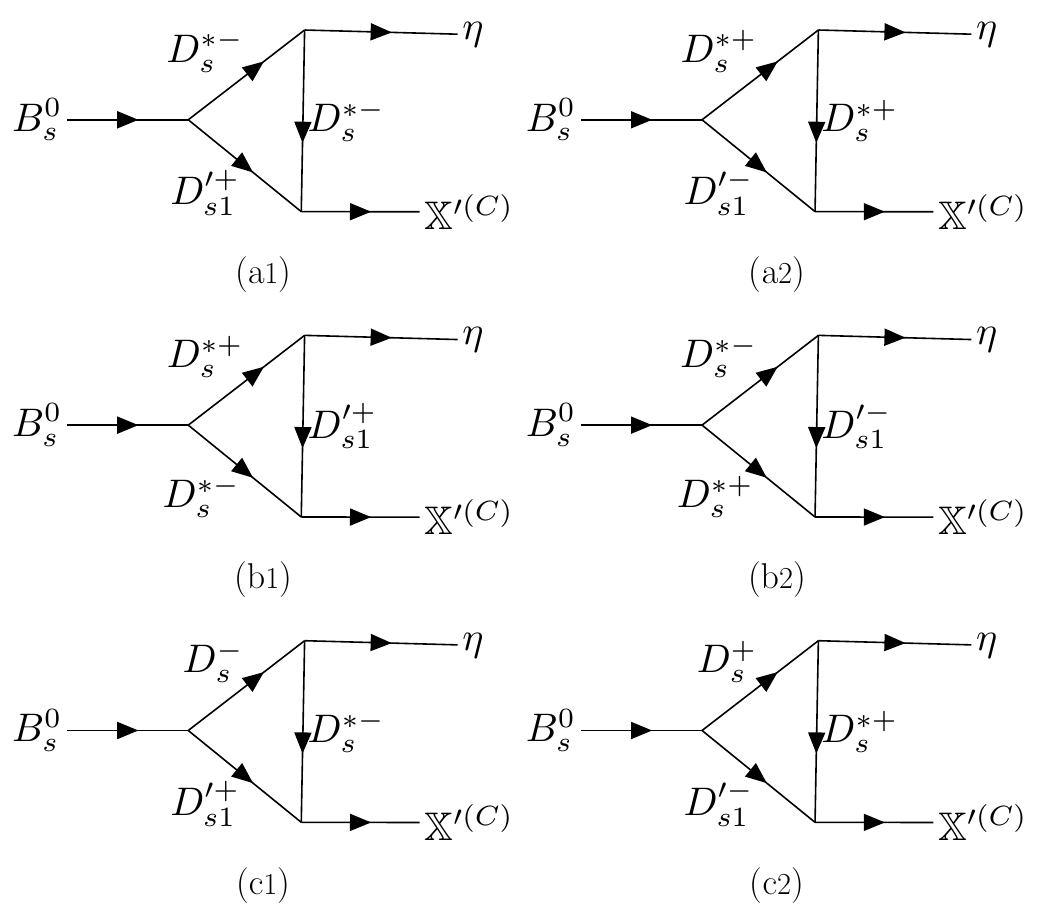}
  \caption{Triangle diagrams for $B_s^0\to\eta\mathbb{X}^{\prime(C)}$.
    Panels (a1), (b1), and (c1) produce $D_{s1}^{\prime+}D_s^{*-}$;
    panels (a2), (b2), and (c2) produce $D_{s1}^{\prime-}D_s^{*+}$.
    For $\widetilde{\mathbb{X}}^{(C)}$, replace $D_{s1}'$ by $D_{s1}$
    and $\mathbb{X}'$ by $\widetilde{\mathbb{X}}$.
    The momenta $r$, $q$, and $\ell$ follow the same convention as in
    Fig.~\ref{fig:Ds0loops}.}
  \label{fig:Ds1loops}
\end{figure}

Figures~\ref{fig:Ds0loops} and \ref{fig:Ds1loops} each show six production diagrams, labeled (a1)--(c2). The letters a, b, and c identify topology pairs. In every pair, digit 1 selects the first molecular charge component and digit 2 the second, in the order of Eq.~\eqref{eq:Xmolwave}; all diagrams have the same initial $B_s^0$. We denote the single-diagram amplitudes before charge projection---that is, before projection onto the molecular charge combinations in Eq.~\eqref{eq:Xmolwave}---by $\mathcal A_{\mathrm{a1}},\ldots,\mathcal A_{\mathrm{c2}}$. 
For the loop diagrams in Figs.~\ref{fig:Ds0loops} and \ref{fig:Ds1loops}, we write $p_0=p_1+p_2$, where $p_1$ and $p_2$ are the $\eta$ and molecular momenta, and route the internal momenta as
\begin{equation}
r=p_1+\ell,\qquad q=p_2-\ell,\qquad r+q=p_0.
\label{eq:routing}
\end{equation}
Here $r$ and $q$ are the momenta of the upper and lower internal lines connected to the $B_s^0$ weak vertex, respectively, while $\ell$ is the momentum of the vertical exchanged line. These labels refer to positions in the triangle diagrams, not to a fixed emission or spectator assignment: the external emitted meson is attached to the upper line in some panels and to the lower line in others. We use $g_{\mu\nu}=\operatorname{diag}(1,-1,-1,-1)$ and $\varepsilon_{0123}=+1$.

\subsection{Weak vertices and form factors}\label{sec:weakvertices}
We model the weak sources by the color-allowed external-$W$ emission shown for $B_s^0\to D_s^-D_s^+$ in Fig.~\ref{fig:weakemission}. The same quark-level diagram generates the vector and positive-parity charmed-strange sources used here. For a transition meson $M_t^-$ and an emitted meson $M_e^+$, we retain this topology in the factorized form
\begin{align}
\mathcal A(B_s^0\to M_t^-M_e^+)&=\mathcal K
\langle M_t^-|\bar b\gamma_\mu(1-\gamma_5)c|B_s^0\rangle
\nonumber\\[-2pt]&\quad\times
\langle M_e^+|\bar c\gamma^\mu(1-\gamma_5)s|0\rangle,\label{eq:factorization}\\
\mathcal K&=\frac{G_F}{\sqrt2}V_{cb}^*V_{cs}a_1^{\rm eff}.\label{eq:weakK}
\end{align}
Here $a_1^{\rm eff}$ collects the color-weighted current-current coefficients for the external-emission contraction. 
We use Eqs.~\eqref{eq:factorization} and \eqref{eq:weakK} as a phenomenological weak-source model. The full $B_s^0\to D_s^-D_s^+$ amplitude also admits $W$-exchange, penguin and penguin-annihilation contributions~\cite{Bel:2015wha}; these are not included in the retained emission source.

We write $m_t\equiv m_{M_t}$ and $m_e\equiv m_{M_e}$. The vacuum currents are
\begin{equation}\begin{aligned}
L_{D_s}^{\mu}&=-if_{D_s}k_e^\mu,&L_{D_{s0}^*}^{\mu}&=f_{D_{s0}^*}k_e^\mu,\\
L_{D_s^*}^{\mu}&=f_{D_s^*}m_{D_s^*}\epsilon_e^{*\mu},&
L_{D_{s1}^{(\prime)}}^{\mu}&=-f_{D_{s1}^{(\prime)}}m_{D_{s1}^{(\prime)}}\epsilon_e^{*\mu}.
\end{aligned}\label{eq:decayconstants}\end{equation}
The transition-current decompositions are listed in Appendix~\ref{app:vertices}. Contracting these currents with Eq.~\eqref{eq:decayconstants} defines the weak tensor $\mathcal W_{M_t,M_e}(k_t,k_e)$, with the external polarizations removed. Its arguments always give the transition momentum first.

For the limited purpose of evaluating the factorized sources, we write the fitted scalar functions as
\begin{equation}
F(t)=\frac{F(0)}{1-a\,t/m_{B_s}^2+b\,(t/m_{B_s}^2)^2}.
\label{eq:ffparam}
\end{equation}
Table~\ref{tab:formfactors} lists the actual central inputs. The $D_{s1}$ and $D_{s1}^\prime$ convention of Ref.~\cite{Yang:2024qij} differs from Eq.~\eqref{eq:particles}; the numerical inputs are therefore matched by physical mass. The auxiliary functions are computed from the same fitted functions.

The analytic current parameterizations satisfy $A_3(0)=A_0(0)$ and $V_3(0)=V_0(0)$, which remove the apparent poles at $t=0$. The fitted parameters do not exactly satisfy these relations: $A_3(0)-A_0(0)=0.00852$, while $V_3(0)-V_0(0)=0.02236$ and $0.00351$ for $D_{s1}(2460)$ and $D_{s1}(2536)$, respectively. We therefore use these fits only as phenomenological inputs at the physical emitted-meson point $t=m_e^2$. They are neither extrapolated to the loop-dependent $k_e^2$ nor interpreted as constrained, complete $t$-dependent currents. No additional kinematic constraint at maximum recoil ($t=0$) is imposed on the tabulated parameters.

\begin{table}[t]
\caption{Weak form-factor parameters from Ref.~\cite{Yang:2024qij}, in the particle convention of Eq.~\eqref{eq:particles}. The form factor $A$ in a given row belongs to the transition shown in the first column. These fits are used only at $t=m_e^2$, as described in the text.}
\label{tab:formfactors}\centering
\begin{tabular}{llrrr}\toprule
Transition & $F$ & $F(0)$ & $a$ & $b$\\\midrule
$B_s\to D_s$ & $F_1$ & 0.65 & 0.84 & 1.02\\
& $F_0$ & 0.65 & 0.50 & 0.99\\
$B_s\to D_{s0}^*$ & $F_1$ & 0.21 & 0.63 & 0.78\\
& $F_0$ & 0.21 & $-0.43$ & 0.28\\
$B_s\to D_s^*$ & $A$ & 0.72 & 0.86 & 1.10\\
& $A_0$ & 0.65 & 0.23 & 0.21\\
& $A_1$ & 0.62 & 0.48 & 0.32\\
& $A_2$ & 0.57 & 0.80 & 0.95\\
$B_s\to D_{s1}'$ & $A$ & 0.20 & $-0.27$ & 0.11\\
& $V_0$ & 0.40 & $-0.17$ & $-0.02$\\
& $V_1$ & 0.58 & $-0.05$ & 0.02\\
& $V_2$ & $-0.05$ & 0.56 & 2.50\\
$B_s\to D_{s1}$ & $A$ & 0.08 & 2.05 & 5.57\\
& $V_0$ & $-0.08$ & 1.24 & 0.74\\
& $V_1$ & 0.17 & $-0.52$ & 0.36\\
& $V_2$ & 0.11 & 0.25 & $-0.07$\\\bottomrule
\end{tabular}\end{table}

Within this phenomenological on-shell prescription, the complete scalar coefficient of each current is evaluated at $t=m_e^2$, while its four-vectors retain their loop dependence. Thus, for example,
\begin{equation}\begin{aligned}
\frac{m_{B_s}^2-m_t^2}{k_e^2}[F_0(k_e^2)-F_1(k_e^2)]
\\\longrightarrow\quad
\frac{m_{B_s}^2-m_t^2}{m_e^2}[F_0(m_e^2)-F_1(m_e^2)].
\end{aligned}\label{eq:freeze}\end{equation}
This prescription is applied to the complete scalar coefficient of every weak current.

\subsection{Strong transitions}\label{sec:strongvertices}
\label{sec:HH}
We describe every retained strong emission vertex using the same heavy-quark and chiral expansion. For the strange heavy mesons, the standard $H,S,T$ superfields can be written directly in terms of the physical fields~\cite{Casalbuoni:1996pg,Colangelo:2012xi}:
\begin{align}
H_s&=\frac{1+\slashed v}{2}(D_{s\mu}^*\gamma^\mu-D_s\gamma_5),\nonumber\\
S_s&=\frac{1+\slashed v}{2}(D_{s1}^{\prime\mu}\gamma_\mu\gamma_5-D_{s0}^*),\nonumber\\
T_s^\mu&=\frac{1+\slashed v}{2}\left\{D_{s2}^{*\mu\nu}\gamma_\nu\right.\nonumber\\
&\quad\left.-\sqrt{\frac32}D_{s1\nu}\gamma_5
\left[g^{\mu\nu}-\frac13\gamma^\nu(\gamma^\mu-v^\mu)\right]\right\}.
\label{eq:superfields}
\end{align}
Here $v^2=1$ and $\bar H=\gamma^0H^\dagger\gamma^0$, with analogous definitions for $S,T$. The superfields contain the $s_\ell^P=\frac12^-$, $\frac12^+$ and $\frac32^+$ doublets, respectively. The two axial states are treated as unmixed heavy-spin eigenstates. The subscripts $a,b$ in the flavor traces below run over light flavors; their strange components are given explicitly above.

With $\xi=\exp(i\Phi/f_\pi)$, define
\begin{align}
\mathcal A_\mu&=\frac{i}{2}(\xi^\dagger\partial_\mu\xi-\xi\partial_\mu\xi^\dagger)
=-\frac{\partial_\mu\Phi}{f_\pi}+O(\Phi^3),\nonumber\\
\mathcal V_\mu&=\frac12(\xi^\dagger\partial_\mu\xi+\xi\partial_\mu\xi^\dagger).
\label{eq:axialcurrent}
\end{align}
The pseudoscalar octet entering $\xi$ is
\begin{equation}
\Phi=\begin{pmatrix}
\frac{\pi^0}{\sqrt2}+\frac{\eta}{\sqrt6} & \pi^+ & K^+\\[2pt]
\pi^- & -\frac{\pi^0}{\sqrt2}+\frac{\eta}{\sqrt6} & K^0\\[2pt]
K^- & \bar K^0 & -\sqrt{\frac23}\eta
\end{pmatrix},
\label{eq:phioctet}
\end{equation}
where $\eta$ denotes the octet field $\eta_8$. The covariant derivative uses the connection $\mathcal V_\mu$. The external $\eta$ is identified with this octet field and assigned its physical mass. The interaction terms required for the original triangle set are
\begin{align}
\mathcal L_{HH}&=g_{\rm HH}\,\operatorname{Tr}(\bar H_aH_b\gamma_\mu\gamma_5\mathcal A^\mu_{ba}),\nonumber\\
\mathcal L_{SH}&=h\,\operatorname{Tr}(\bar H_aS_b\gamma_\mu\gamma_5\mathcal A^\mu_{ba})+\hc,\nonumber\\
\mathcal L_{TH}&=\frac{h'}{\Lambda_\chi}\operatorname{Tr}\left[\bar H_aT_b^\mu
(iD_\mu\slashed{\mathcal A}+i\slashed D\mathcal A_\mu)_{ba}\gamma_5\right]+\hc.
\label{eq:HHlag}
\end{align}
Charge-conjugated heavy fields provide the anticharmed vertices; their phases are fixed by Eq.~\eqref{eq:Cphases}. The operators proportional to $h$ and $h'$ describe $S$- and $D$-wave emission, respectively. Heavy-quark symmetry relates the spin partners within each operator.

The numerical inputs are $g_{\rm HH}=0.59$, $h=0.56$, $h'=0.43$, $f_\pi=0.132$ GeV and $\Lambda_\chi=1$ GeV. The $h,h'$ conventions follow Ref.~\cite{Colangelo:2012xi}, and $g_{\rm HH}$ follows Ref.~\cite{Isola:2003fh}. Equations~\eqref{eq:HHlag} generate the $D_sD_s^*\eta$ and $D_s^*D_s^*\eta$ transitions, the $S$-wave $D_{s0}^*D_s\eta$ and $D_{s1}'D_s^*\eta$ transitions, and the $D$-wave $D_{s1}D_s^*\eta$ transition. The transition tensors are collected in Appendix~\ref{app:strongmol}.

\subsection{Molecular vertices and constituent couplings}\label{sec:molecular}\label{sec:compositeness}
The constituent couplings refer to normalized spin and charge channels. With the dimensionless partial-wave convention
\begin{equation}\begin{aligned}
T_{ij}(s)&=-\frac{g_i g_j}{s-s_R}+\text{regular terms},\\
s_R&=(M_R-i\Gamma_R/2)^2,
\end{aligned}\label{eq:rescomp}\end{equation}
the couplings $g_i$ have dimension one. The complete molecular--constituent interactions, including both charge components, are
\begin{align}
\mathcal L_{\mathbb X}&=\frac{g_{\mathbb X}}{\sqrt2}\,
\mathbb X^{(C)\mu}\left(D_{s0}^{*+}D_{s\mu}^{*-}
-C D_{s0}^{*-}D_{s\mu}^{*+}\right)+\hc,\nonumber\\
\mathcal L_{\mathbb X'}&=\frac{g_{\mathbb X'}}{2M_R}\,
\varepsilon_{\mu\nu\alpha\beta}
(\partial^\mu\mathbb X^{\prime(C)\nu})\nonumber\\
&\quad\times\left(D_{s1}^{\prime+\alpha}D_s^{*-\beta}
-C D_{s1}^{\prime-\alpha}D_s^{*+\beta}\right)+\hc,\nonumber\\
\mathcal L_{\widetilde{\mathbb X}}&=\frac{g_{\widetilde{\mathbb X}}}{2M_R}\,
\varepsilon_{\mu\nu\alpha\beta}
(\partial^\mu\widetilde{\mathbb X}^{(C)\nu})\nonumber\\
&\quad\times\left(D_{s1}^{+\alpha}D_s^{*-\beta}
-C D_{s1}^{-\alpha}D_s^{*+\beta}\right)+\hc .
\label{eq:mollag}
\end{align}
Here $C=\pm1$ labels the two charge-conjugation combinations in Eq.~\eqref{eq:Xmolwave}, and the derivative acts on the molecular field. The axial--vector interactions contain both the charge normalization $1/\sqrt2$ and the spin normalization $1/\sqrt2$ used below. Removing only the charge coefficient in Eq.~\eqref{eq:Xmolwave}, the molecular tensors are
\begin{align}
U_{\mathbb{X},\nu}&=g_{\mathbb{X}}\epsilon_{2\nu}^*,\nonumber\\
U_{\mathbb{X}',\alpha\beta}&=\frac{g_{\mathbb{X}'}}{\sqrt2 M_R}
\varepsilon_{\mu\nu\alpha\beta}p_2^\mu\epsilon_2^{*\nu},\nonumber\\
U_{\widetilde{\mathbb{X}},\alpha\beta}&=\frac{g_{\widetilde{\mathbb{X}}}}{\sqrt2 M_R}
\varepsilon_{\mu\nu\alpha\beta}p_2^\mu\epsilon_2^{*\nu}.
\label{eq:Umol}
\end{align}
The indices in the last two lines refer to the axial and vector constituents, respectively. The factor $1/\sqrt2$ in these tensors normalizes their spin-one coupling, separately from the charge normalization. 

For above-threshold assignments we estimate the coupling magnitudes using the two-body loop function $G_i(s)$ and the width-saturation construction of Refs.~\cite{Guo:2015daa,Guo:2019kdc,Guo:2020vmu}. For a channel with constituent masses $m_1,m_2$, our scalar-loop normalization is, up to an irrelevant subtraction constant,
\begin{equation}\begin{aligned}
G_i(s)&=\frac1{16\pi^2}\int_0^1\dd x\,\ln\frac{D_i(x,s)}{\mu_0^2},\\
D_i(x,s)&=x m_1^2+(1-x)m_2^2-x(1-x)s,\\
G_i'(s)&\equiv\frac{\dd G_i}{\dd s}
=-\frac1{16\pi^2}\int_0^1\dd x\,\frac{x(1-x)}{D_i(x,s)} .
\end{aligned}\label{eq:Gdefinition}\end{equation}
Here $\mu_0$ is a subtraction scale; only the subtraction-independent derivative enters below. The required residue relations are
\begin{equation}\begin{aligned}
X_i&=-g_i^2 G_i'(s_R),\\
\widehat X_i&=|g_i|^2|G_i'(s_R)|,\qquad
\widehat X=\sum_i\widehat X_i.
\end{aligned}\label{eq:positive}\end{equation}
For a resonance, $X_i$ is generally complex~\cite{Kinugawa:2024crb}, whereas $\widehat X_i$ is a deliberately constructed positive real estimator rather than a probability equal to $X_i$. For each above-threshold coupling calibration, we use two channels: the selected constituent channel and $h_c\eta$ for $C=-1$, or $\chi_{c1}\eta$ for $C=+1$. Writing $d_i=|G_i'(s_R)|$ and $\Gamma_i=w_i|g_i|^2$, where $\Gamma_i$ is the partial width of channel $i$, the couplings follow from
\begin{equation}
\begin{pmatrix}d_{\rm mol}&d_{\rm low}\\w_{\rm mol}&w_{\rm low}\end{pmatrix}
\begin{pmatrix}|g_{\rm mol}|^2\\|g_{\rm low}|^2\end{pmatrix}
=\begin{pmatrix}\widehat X\\\Gamma_R\end{pmatrix}.
\label{eq:saturation}
\end{equation}
The near-threshold width coefficient is
\begin{equation}\begin{aligned}
w_{\rm mol}&=\int_{m_1+m_2}^{M_R+10\Gamma_R}\!\dd E\,
\frac{k(E^2)}{8\pi E^2}
\frac{\Gamma_R/(2\pi)}{(E-M_R)^2+\Gamma_R^2/4},\\
w_{\rm low}&=\frac{k_{\rm low}(M_R^2)}{8\pi M_R^2},\qquad
k(s)=\frac{\lambda^{1/2}(s,m_1^2,m_2^2)}{2\sqrt s}.
\end{aligned}\label{eq:widthkernel}\end{equation}
Here $\lambda$ is the K\"all\'en function. In these conventions $g_i$ is expressed in GeV, $d_i$ in GeV$^{-2}$, and $w_i$ in GeV$^{-1}$. For channels open at the central mass we use the second-sheet continuation
\begin{equation}\begin{aligned}
G_{\rm II}(s)&=G_{\rm I}(s)+\frac{i k_{\rm phys}(s)}{4\pi\sqrt s},\\
G_{\rm II}'(s)&=G_{\rm I}'(s)+\frac{i}{4\pi}\frac{\dd}{\dd s}
\frac{k_{\rm phys}(s)}{\sqrt s},
\end{aligned}\label{eq:sheetII}\end{equation}
with $\operatorname{Im}k_{\rm phys}\geq0$ and $k_{\rm phys}$ evaluated from Eq.~\eqref{eq:widthkernel}. Closed channels use sheet I. This specifies the branch convention used in Eq.~\eqref{eq:saturation}.

For the above-threshold assignments, the central compositeness input is chosen from the effective-range estimate (ERE)~\cite{Gao:2018jhk},
\begin{equation}\begin{aligned}
\widehat X_{\rm ERE}&=\left|\frac{\operatorname{Im}k_R}{\operatorname{Re}k_R}\right|,\\
k_R&=\sqrt{2\mu\bigl(M_R-m_1-m_2-i\Gamma_R/2\bigr)},
\end{aligned}\label{eq:ERE}\end{equation}
where $\mu=m_1m_2/(m_1+m_2)$. Although $\widehat X_{\rm ERE}$ is estimated from the selected molecular channel, we adopt the model assumption $\widehat X=\widehat X_{\rm ERE}$ for the total input in the two-channel equation~\eqref{eq:saturation}; this is not an identity between one-channel and total compositeness.

For an assignment whose central mass lies below the selected molecular threshold, we use instead the real-bound-state Weinberg relation~\cite{Weinberg:1965hj}. With $M_B=M_R$, define
\begin{equation}
B=m_1+m_2-M_B>0,\qquad \kappa=\sqrt{2\mu B}.
\label{eq:binding}\end{equation}
On sheet I at the real bound-state point, the weak-binding approximation gives
\begin{equation}
|g_{\rm mol}|^2=-\frac{X_{\rm mol}}{G_{\rm I}'(M_B^2)}
\simeq \frac{16\pi M_B^2\kappa}{\mu}X_{\rm mol},
\label{eq:weinberg}\end{equation}
and we set $X_{\rm mol}=1$. Here $X_{\rm mol}$ is the real molecular weight of the bound state, not the total resonance estimator $\widehat X$ in Eq.~\eqref{eq:saturation}. The experimental width does not enter Eq.~\eqref{eq:weinberg}; accordingly, this branch has no width-saturation step, no independently determined $g_{\rm low}$, and no width-error propagation. Thus the calculation rule is fixed by the central mass relative to the selected threshold: above threshold we use Eqs.~\eqref{eq:saturation} and \eqref{eq:ERE}, while below threshold we use Eq.~\eqref{eq:weinberg} with $X_{\rm mol}=1$. The five resulting calibrations and the opposite-$C$ partner prescription are given in Sec.~\ref{sec:inputs}.

\subsection{Triangle amplitudes and relative phases}\label{sec:propagator}

A spin-one internal line is represented by
\begin{equation}
\Delta_i^{\mu\nu}(k)=\frac{i\Pi^{\mu\nu}(v)}{k^2-m_i^2+i0}
\label{eq:propagators}
\end{equation}
with $\Pi^{\mu\nu}(v)=-g^{\mu\nu}+v^\mu v^\nu$, and a spin-zero line by $i/(k^2-m_i^2+i0)$. The denominators retain the physical masses and relativistic momentum dependence.
The calculation is performed in the static limit with
\begin{equation}
v^\mu=\frac{p_0^\mu}{m_{B_s}}=(1,\mathbf0).
\label{eq:velocity}
\end{equation}

Each single-diagram amplitude before charge projection has the common form
\begin{equation}
\mathcal A_j=\mathcal I_{12,3}[N_j]
=i\int\frac{\dd^4\ell}{(2\pi)^4}
\frac{N_j(r,q,\ell)\mathcal F_1(\ell^2)}{D_1(r)D_2(q)D_3(\ell)},
\label{eq:loopgeneric}
\end{equation}
where $D_i(k)=k^2-m_i^2+i0$. The numerator contracts the weak tensor, strong-transition tensor, molecular vertex, and the spin projectors. It includes $g_{\rm mol}$ but not the molecular charge-projection coefficients. The individual contractions are given in Appendix~\ref{app:loops}.

To account for the off-shell effects of the exchanged state and to regularize the ultraviolet behavior of the loop integral, we introduce a monopole form factor~\cite{Gortchakov:1995im,Cheng:2004ru,Wu:2019vbk,Guo:2010ak,Yu:2024mol},
\begin{equation}\begin{aligned}
\mathcal F_1(\ell^2)&=\frac{\Lambda^2-m_{\rm ex}^2}{\Lambda^2-\ell^2-i0},\\
\Lambda&=m_{\rm ex}+\alpha\Lambda_{\rm QCD},\qquad
\Lambda_{\rm QCD}=0.220~\mathrm{GeV}.
\end{aligned}\label{eq:monopole}\end{equation}
The retained numerators have rank at most three, for which this regulator makes the triangle integrals ultraviolet finite. We evaluate them using
\begin{equation}
\frac{\mathcal F_1(\ell^2)}{\ell^2-m_{\rm ex}^2+i0}
=\frac1{\ell^2-m_{\rm ex}^2+i0}-\frac1{\ell^2-\Lambda^2+i0}.
\label{eq:PVidentity}
\end{equation}
Only the exchanged denominator mass changes in the subtraction; the vertices remain fixed.

With the charge ordering of Figs.~\ref{fig:Ds0loops} and~\ref{fig:Ds1loops}, the coherent charge projections are
\begin{align}
\mathcal M_{\mathrm a}&=\frac{\mathcal A_{\mathrm{a1}}-Cs_{\mathrm a}\mathcal A_{\mathrm{a2}}}{\sqrt2},
\quad s_{\mathrm a}=e^{i\delta_{\mathrm a}}=\pm1,\nonumber\\
\mathcal M_{\mathrm b}&=\frac{\mathcal A_{\mathrm{b1}}-C\mathcal A_{\mathrm{b2}}}{\sqrt2},\nonumber\\
\mathcal M_C&=\mathcal M_{\mathrm a}+e^{i\theta_{\mathrm b}}\mathcal M_{\mathrm b}.
\label{eq:phaseamp}
\end{align}
For the retained tensors $\mathcal A_{\mathrm{c1}}=\mathcal A_{\mathrm{c2}}=0$. The relative source phase between diagrams (b1) and (b2) is fixed by the common $H$-doublet convention in the factorized scalar-channel sources, and by their common $D_s^{*+}D_s^{*-}$ weak source for the axial channels. The two $s_{\mathrm a}$ choices describe the remaining relative source sign, independently of the molecular $C$ parity.

For a general source phase, the branching fraction and normalization are
\begin{equation}
\mathcal B_C=\mathcal N|\mathcal M_C|^2,
\qquad
\mathcal N=\frac{\tau_{B_s}}{\hbar}\frac{p_\eta}{8\pi m_{B_s}^2},
\label{eq:generalrate}
\end{equation}
where $p_\eta=\lambda^{1/2}(m_{B_s}^2,m_\eta^2,M_R^2)/(2m_{B_s})$. Here $\mathbb X$ is treated as an on-shell spin-one state of fixed mass $M_R$, which fixes $p_\eta$ and its longitudinal polarization. Neither spectral-function integration nor nonresonant amplitudes with the same hadronic content are included; $\mathcal B_C$ is the resulting resonant $B_s^0\to\eta\mathbb X$ branching fraction. In the numerical reference we fix the relative phase of the complete amplitudes $\mathcal M_{\mathrm a}$ and $\mathcal M_{\mathrm b}$,
\begin{equation}
\phi_{\mathrm{ab}}\equiv\arg\!\left(e^{i\theta_{\mathrm b}}\mathcal M_{\mathrm b}\right)-\arg\mathcal M_{\mathrm a}=\frac\pi2
\quad(\mathcal M_{\mathrm b}\neq0).
\label{eq:phiab}
\end{equation}
The corresponding branching fraction is
\begin{equation}
\mathcal B_C^{(\pi/2)}=\mathcal N\bigl(|\mathcal M_{\mathrm a}|^2+|\mathcal M_{\mathrm b}|^2\bigr).
\label{eq:productionwidth}
\end{equation}
This condition is not generally equivalent to fixing the additional source phase $\theta_{\mathrm b}=\pi/2$. Instead, $\theta_{\mathrm b}=\pi/2+\arg\mathcal M_{\mathrm a}-\arg\mathcal M_{\mathrm b}$ modulo $2\pi$, and it can vary with $\alpha$ and $s_{\mathrm a}$. 
Interference within each pair is retained.

\subsection{Numerical inputs and coupling uncertainties}\label{sec:inputs}
We use the following meson masses in GeV~\cite{ParticleDataGroup:2024cfk}:
\begin{equation}\begin{aligned}
m_{B_s}&=5.36688,&m_\eta&=0.547862,\\
m_{D_s}&=1.96835,&m_{D_s^*}&=2.11220,\\
m_{D_{s0}^*}&=2.31770,&m_{D_{s1}'}&=2.45950,\\
m_{D_{s1}}&=2.53511,&m_{h_c}&=3.52538,\\
m_{\chi_{c1}}&=3.51067.&&
\end{aligned}\label{eq:massinputs}\end{equation}
The weak inputs are $G_F=1.16638\times10^{-5}$ GeV$^{-2}$, $|V_{cb}|=0.0408$, $|V_{cs}|=0.975$, $a_1^{\rm eff}=1.018$, and $\tau_{B_s}=1.520\times10^{-12}$ s. The decay constants in GeV are $f_{D_s}=0.2575$, $f_{D_s^*}=0.272$, $f_{D_{s0}^*}=0.067$, $f_{D_{s1}'}=0.145$ and $f_{D_{s1}}=0.032$~\cite{Yang:2024qij}. We use $\hbar=6.582119569\times10^{-25}$ GeV\,s.

The experimental mass and width inputs are listed in Table~\ref{tab:poleinputs}. For the benchmarks carrying our $Y(4626)$ model label, we use the PDG 2026 weighted averages for $\psi(4660)$, $M_R=4623\pm10$ MeV and $\Gamma_R=55\pm9$ MeV~\cite{ParticleDataGroup:2026mpi}, motivated by the measurements involving $D_{s1}(2536)$ included in that listing. These are not experimental averages for the Belle $Y(4626)$ signal itself; $Y(4626)$ is retained only as the model label used in our assignment names.

The five experimentally anchored assignments are $\mathbb{X}^{(-)}$ at the $Y(4500)$ mass, $\mathbb{X}^{\prime(+)}$ and $\widetilde{\mathbb{X}}^{(+)}$ at $X(4630)$, and $\mathbb{X}^{\prime(-)}$ and $\widetilde{\mathbb{X}}^{(-)}$ at the $Y(4626)$ label. Different constituent assignments for the same benchmark state are treated as alternative hypotheses. Their thresholds are 4.42990, 4.57170 and 4.64731 GeV for $\mathbb{X}$, $\mathbb{X}'$ and $\widetilde{\mathbb{X}}$, respectively. Thus ERE plus width saturation is applied to $\mathbb{X}^{(-)}[Y(4500)]$ and to the two $\mathbb{X}'$ assignments, whereas Weinberg's relation with $X_{\rm mol}=1$ is applied to both $\widetilde{\mathbb X}$ assignments. With $\Delta=M_R-m_1-m_2$, the three ERE cases have $\Gamma_R/\Delta=2.03$, $3.20$ and $1.07$, respectively; their ERE inputs are therefore exploratory single-channel estimates, not controlled effective-range fits. Table~\ref{tab:couplings} gives the five independent molecular-coupling calibrations.

For each experimentally anchored assignment $\mathsf p$, we additionally consider the opposite-$C$ combination $\mathsf q$ of the same constituents. Our minimal partner prescription is
\begin{equation}
C_{\mathsf q}=-C_{\mathsf p},\qquad
(M_R,\Gamma_R,|g_{\rm mol}|)_{\mathsf q}
=(M_R,\Gamma_R,|g_{\rm mol}|)_{\mathsf p}.
\label{eq:Cpartner}
\end{equation}
Thus the ten production hypotheses consist of five experimentally anchored reference assignments and five opposite-$C$ partner hypotheses. Since $B_s^0$ is not a $C$-parity eigenstate and its weak decay does not conserve $C$, producing either combination is not forbidden by a $C$-parity selection rule. The quantum numbers of an observed resonance constrain the corresponding reference assignment, but do not by themselves exclude a molecular partner with the opposite $C$.

The degeneracy and equal molecular residue in Eq.~\eqref{eq:Cpartner} are additional spectrum assumptions, not consequences of charge conjugation. We adopt them to obtain a controlled production estimate without introducing unconstrained new parameters. No new calibration is performed for a partner: its production calculation reuses the pre-charge-projection loop amplitudes and $|g_{\rm mol}|$ of the corresponding reference assignment, and changes only $C$ in Eq.~\eqref{eq:phaseamp}. For a Weinberg-calibrated reference assignment, only this molecular coupling is inherited; no companion-channel coupling is constructed. Partner rates are therefore conditional estimates rather than predictions of a uniquely determined molecular spectrum.

\begin{table}[t]\centering
\caption{Mass and width inputs in MeV. For $Y(4500)$ and $X(4630)$, the first and second uncertainties are statistical and systematic, respectively. The last row uses the PDG 2026 $\psi(4660)$ weighted averages as numerical inputs for our $Y(4626)$ model label.}
\label{tab:poleinputs}
\small
\setlength{\tabcolsep}{1.5pt}
\resizebox{\columnwidth}{!}{%
\begin{tabular}{lrrl}\toprule
State & $M_R$ & $\Gamma_R$ & Reference\\\midrule
$Y(4500)$ & $4484.7\pm13.3\pm24.1$ & $111.1\pm30.1\pm15.2$ & \cite{BESIII:2022joj}\\
$X(4630)$ & $4626\pm16{}_{-110}^{+18}$ & $174\pm27{}_{-73}^{+134}$ & \cite{LHCb:2021uow}\\
$Y(4626)$ label & $4623\pm10$ & $55\pm9$ & \cite{ParticleDataGroup:2026mpi}\\\bottomrule
\end{tabular}}
\end{table}

\begin{table}[t]\centering
\caption{Molecular couplings $|g_{\rm mol}|$ for the five experimentally anchored assignments. The Coupling column identifies the corresponding constant from Eq.~\eqref{eq:mollag}, and Value gives its magnitude in GeV. ``ERE'' means Eq.~\eqref{eq:ERE} followed by the two-channel width saturation in Eq.~\eqref{eq:saturation}; ``Weinberg'' means Eq.~\eqref{eq:weinberg} with $X_{\rm mol}=1$. The ERE ranges are nonnegative mass--width feasible-domain envelopes, while the Weinberg ranges come from mass variation only. They are not fitted statistical errors. An opposite-$C$ partner inherits $|g_{\rm mol}|$ from the corresponding reference assignment.}
\label{tab:couplings}
\small
\setlength{\tabcolsep}{2.5pt}
\begin{tabular}{lllr}\toprule Assignment & Rule & Coupling & Value [GeV]\\\midrule
$\mathbb{X}^{(-)}\,[Y(4500)]$ & ERE & $|g_{\mathbb X}|$ & $11.79_{-1.45}^{+1.21}$\\
$\mathbb{X}^{\prime(+)}\,[X(4630)]$ & ERE & $|g_{\mathbb X'}|$ & $14.62_{-0.78}^{+0.69}$\\
$\widetilde{\mathbb{X}}^{(+)}\,[X(4630)]$ & Weinberg & $|g_{\widetilde{\mathbb X}}|$ & $14.38_{-4.19}^{+2.10}$\\
$\mathbb{X}^{\prime(-)}\,[Y(4626)]$ & ERE & $|g_{\mathbb X'}|$ & $8.98_{-0.47}^{+0.44}$\\
$\widetilde{\mathbb{X}}^{(-)}\,[Y(4626)]$ & Weinberg & $|g_{\widetilde{\mathbb X}}|$ & $14.86_{-1.82}^{+1.30}$\\
\bottomrule
\end{tabular}

\end{table}

For the ERE cases, the experimental Breit--Wigner mass and width parameters are used as approximate complex-pole inputs. For the Weinberg cases, $M_R$ is instead the real bound-state mass in Eq.~\eqref{eq:weinberg}. The $Y(4500)$ statistical and systematic errors on each side are combined in quadrature, and the published PDG average errors are used directly for the $Y(4626)$ label. For $X(4630)$, only the statistical errors are propagated, giving $M_R=4610$--$4642$ MeV and $\Gamma_R=147$--$201$ MeV. Its systematic errors remain in Table~\ref{tab:poleinputs}, but are not included in the numerical coupling bands: combining the statistical and systematic components in quadrature would raise the upper mass endpoint to about $4650$ MeV, above the $D_{s1}(2536)\bar D_s^*$ threshold and hence outside the domain of either fixed calibration rule adopted here.

For each ERE-calibrated reference assignment, we vary $M_R$ and $\Gamma_R$ over their experimental input intervals while keeping the central $\widehat X$ benchmark, companion channel, and pole-sheet assignment fixed. Solving Eq.~\eqref{eq:saturation}, we retain only solutions with nonnegative $|g_{\rm mol}|^2$ and $|g_{\rm low}|^2$, and use the resulting extrema to define the feasible-domain coupling envelopes. All retained mass intervals remain above the corresponding molecular thresholds. For each Weinberg-calibrated reference assignment, only $M_R$ is varied, with $X_{\rm mol}=1$ and the sheet-I relation held fixed, and the endpoints of the mass interval determine the coupling range. Both mass intervals remain below the $D_{s1}(2536)\bar D_s^*$ threshold. Since $\Gamma_R$ does not enter Eq.~\eqref{eq:weinberg}, its experimental uncertainty does not affect these two bands.


In the production amplitude, all masses, loop functions, weak vertices and strong couplings remain at their central values. Thus the bands propagate only the induced uncertainty in $g_{\rm mol}$:
\begin{equation}
\mathcal B_{\rm lower,upper}(\alpha)
=\mathcal B_{\rm central}(\alpha)
\frac{|g_{\rm mol}|_{\rm min,max}^2}{|g_{\rm mol}|_{\rm central}^2}.
\label{eq:uncertaintyband}
\end{equation}
The same coupling multiplies every diagram of a given assignment, so its uncertainty is fully correlated across the diagrams and throughout the cutoff scan. Each opposite-$C$ partner inherits the dimensionless interval $|g_{\rm mol}|_{\rm min,max}^2/|g_{\rm mol}|_{\rm central}^2$ of its reference assignment; no separate coupling-error scan is constructed for the partner. These shaded bands are conditional input-variation envelopes, not joint $68\%$ confidence intervals. Our numerical results retain some model dependence, and these bands should not be interpreted as a complete quantification of the theoretical uncertainty.

\section{Results and discussion}\label{sec:results}
\subsection{Individual triangle contributions}
\begin{figure*}[!t]\centering
\includegraphics[width=.82\textwidth]{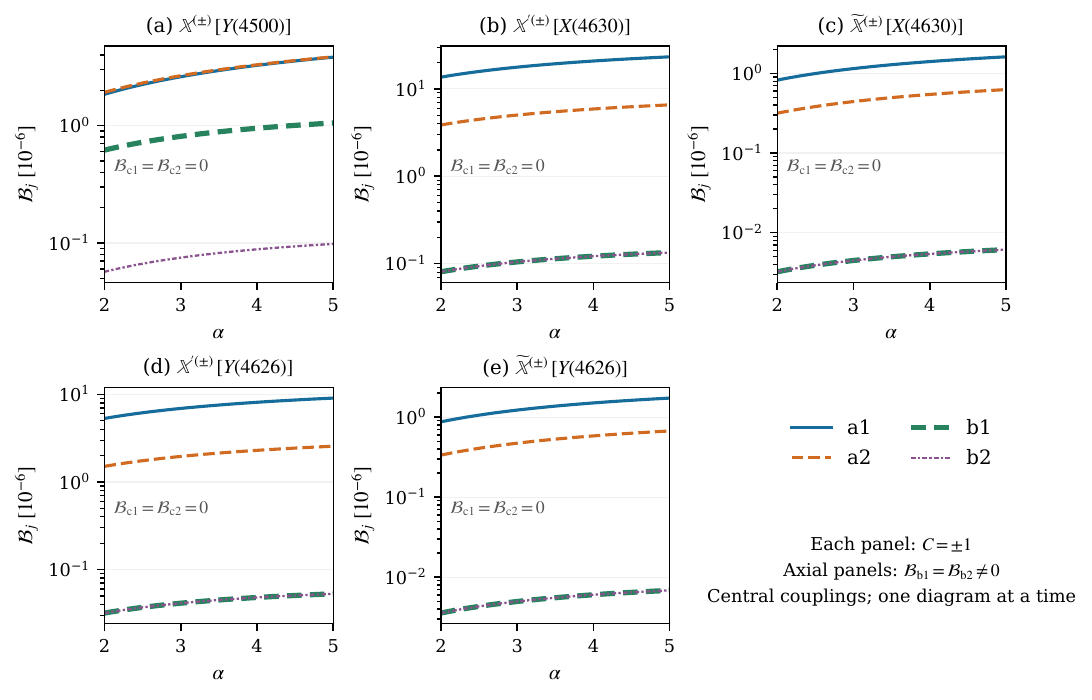}
\caption{Single-diagram branching fractions with central molecular couplings, defined by Eq.~\eqref{eq:diagramdiag} including the charge factor $1/2$. Diagram labels follow Figs.~\ref{fig:Ds0loops} and~\ref{fig:Ds1loops}. Each panel covers both $C=\pm1$ partners; panel (a) includes $\mathbb{X}^{(-)}[Y(4500)]$ and $\mathbb{X}^{(+)}[Y(4500)]$. Their curves coincide before coherent charge projection. The (c1) and (c2) contributions vanish. For axial assignments, the nonzero (b1) and (b2) curves coincide and are shown with different line widths.}
\label{fig:diagranking}\end{figure*}

Figure~\ref{fig:diagranking} compares the cutoff dependence of the individual diagrams using the central molecular couplings. For each panel contribution, we plot
\begin{equation}
\mathcal B_j=\frac{\mathcal N}{2}|\mathcal A_j|^2,
\quad j\in\{\mathrm{a1},\mathrm{a2},\mathrm{b1},\mathrm{b2},\mathrm{c1},\mathrm{c2}\},
\label{eq:diagramdiag}
\end{equation}
including the charge coefficient but retaining only that diagram. The scan covers $2\leq\alpha\leq5$.

For $\mathbb{X}^{(-)}[Y(4500)]$, diagrams (a1) and (a2) are comparable and largest, followed by (b1) and (b2). At $\alpha=4$, their shares of the diagonal sum are approximately $42.9\%$, $43.4\%$, $12.5\%$ and $1.2\%$. The near equality of the (a1) and (a2) contributions is a numerical consequence of their common loop structure and the chosen weak source parameters, not an exact symmetry relation. For the axial assignments, diagram (a1) dominates: the (a1) and (a2) shares are about $77.3\%,21.8\%$ for $\mathbb{X}'$ and $71.7\%,27.7\%$ for $\widetilde{\mathbb{X}}$. Each diagram in the secondary topology class contributes less than $0.5\%$ of the diagonal sum. The ordering is stable over the plotted cutoff range.

The contractions for diagrams (c1) and (c2) vanish for the retained tensors. With $v=p_0/m_{B_s}$, their single antisymmetric contraction has insufficient independent external directions to form the spin-zero decay amplitude. Because Eq.~\eqref{eq:Cpartner} reuses both the amplitudes before charge projection and the molecular coupling, the individual curves in Fig.~\ref{fig:diagranking} are unchanged for an opposite-$C$ partner. Differences between the rates of a reference assignment and its partner arise only from the coherent projections in Eq.~\eqref{eq:phaseamp}.

For the axial assignments, $\mathcal A_{\mathrm{b1}}=-\mathcal A_{\mathrm{b2}}\neq0$, so the individual contributions obey $\mathcal B_{\mathrm{b1}}=\mathcal B_{\mathrm{b2}}\neq0$ and completely overlap in Fig.~\ref{fig:diagranking}(b--e). Their cancellation occurs only after the coherent $C=-1$ projection, $\mathcal M_{\mathrm b}=0$; for $C=+1$ the two amplitudes add. Independently of the constituent type, Eq.~\eqref{eq:phaseamp} implies
\begin{equation}
\mathcal M_{\mathrm a}^{(C)}(s_{\mathrm a})=\mathcal M_{\mathrm a}^{(-C)}(-s_{\mathrm a}),
\label{eq:Csignexchange}
\end{equation}
so reversing $C$ interchanges the two sign choices $s_{\mathrm a}=\pm1$ associated with the leading source pair. These relations explain why the coherent rates can differ substantially from the sum of the individual curves and why reference--partner totals need not be related by a sign exchange alone.

\subsection{Branching fractions and coupling uncertainty bands}
\begin{figure*}[!t]\centering
\includegraphics[width=.92\textwidth]{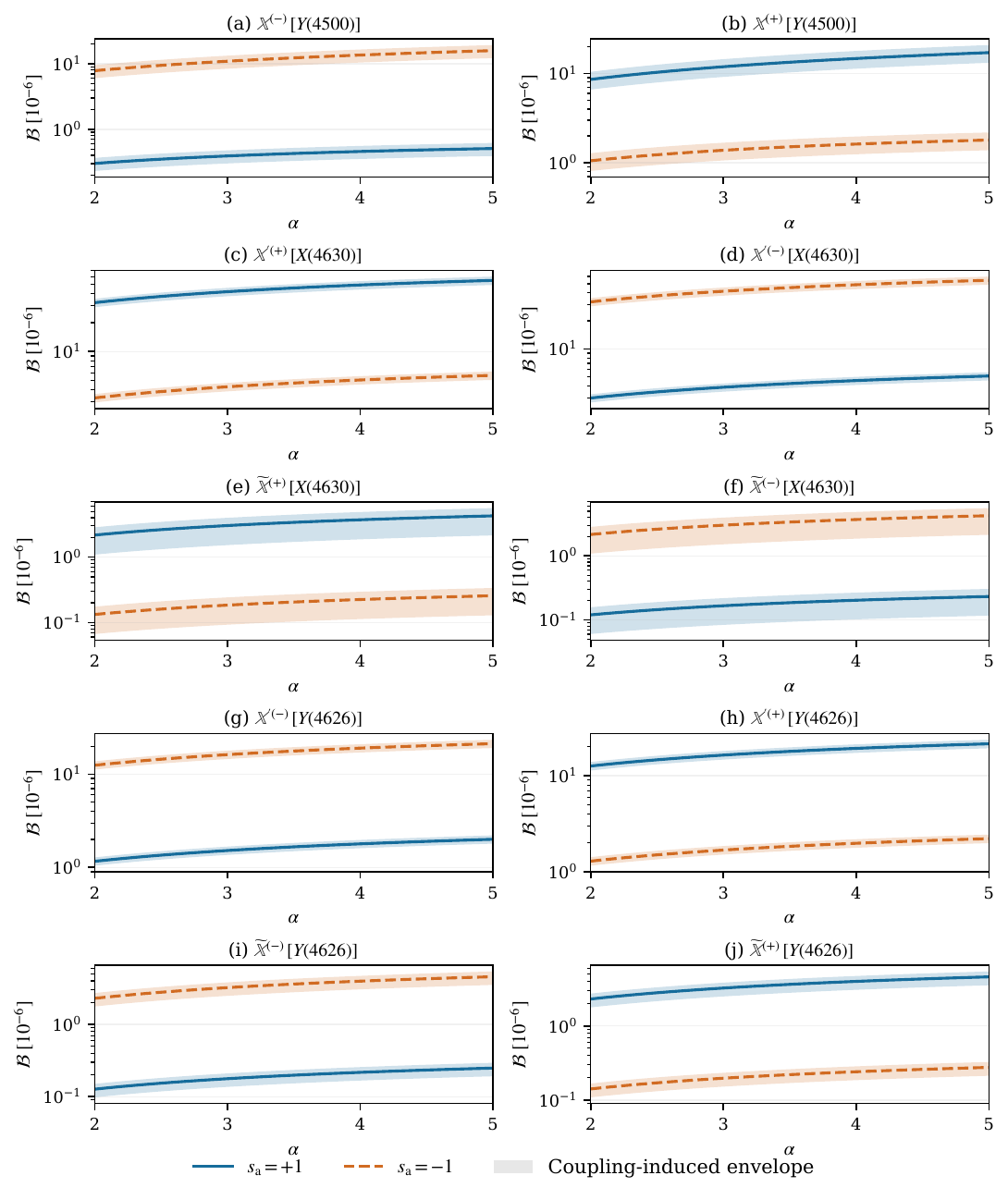}
\caption{Charge-projected branching fractions versus $\alpha$ for all ten hypotheses. Each reference assignment is followed by its opposite-$C$ partner. Solid blue and dashed orange curves correspond to $s_{\mathrm a}=+1$ and $-1$, respectively, with $\mathcal B=\mathcal B_{\mathrm a}+\mathcal B_{\mathrm b}$. For the ERE-calibrated reference assignments, the bands are nonnegative mass--width feasible-domain envelopes; for the Weinberg-calibrated reference assignments, they use mass variation only. Partners inherit the same dimensionless coupling interval. These conditional envelopes are not joint confidence intervals.}
\label{fig:rates}\end{figure*}

Figure~\ref{fig:rates} shows the physical charge-projected rates for $s_{\mathrm a}=+1$ and $-1$. The complete amplitudes of the leading and secondary topology classes, $\mathcal M_{\mathrm a}$ and $\mathcal M_{\mathrm b}$, have relative phase $\pi/2$, as in Eq.~\eqref{eq:productionwidth}. The shading is the molecular-coupling envelope of Eq.~\eqref{eq:uncertaintyband}; the two source-sign choices are displayed as separate curves. The central values at selected cutoffs are collected in Table~\ref{tab:BR4630}.

\begin{table*}[!t]\centering
\caption{Central branching fractions in units of $10^{-6}$, with the complete amplitudes of the leading and secondary topology classes at relative phase $\pi/2$. Each experimentally anchored assignment is immediately followed by its opposite-$C$ partner, evaluated with the inherited coupling prescription of Eq.~\eqref{eq:Cpartner}.}
\label{tab:BR4630}
\begin{tabular}{lrrrrr}\toprule Assignment & $s_{\mathrm a}$ & $\alpha=2$ & $3$ & $4$ & $5$\\\midrule
$\mathbb{X}^{(-)}\,[Y(4500)]$ & $+1$ & 0.301 & 0.392 & 0.459 & 0.509\\
$\mathbb{X}^{(-)}\,[Y(4500)]$ & $-1$ & 7.84 & 10.9 & 13.6 & 15.9\\
$\mathbb{X}^{(+)}\,[Y(4500)]$ & $+1$ & 8.59 & 11.9 & 14.7 & 17.2\\
$\mathbb{X}^{(+)}\,[Y(4500)]$ & $-1$ & 1.05 & 1.38 & 1.62 & 1.79\\
$\mathbb{X}^{\prime(+)}\,[X(4630)]$ & $+1$ & 32.3 & 42.0 & 49.3 & 55.0\\
$\mathbb{X}^{\prime(+)}\,[X(4630)]$ & $-1$ & 3.30 & 4.30 & 5.07 & 5.66\\
$\mathbb{X}^{\prime(-)}\,[X(4630)]$ & $+1$ & 2.98 & 3.88 & 4.58 & 5.13\\
$\mathbb{X}^{\prime(-)}\,[X(4630)]$ & $-1$ & 32.0 & 41.5 & 48.8 & 54.5\\
$\widetilde{\mathbb{X}}^{(+)}\,[X(4630)]$ & $+1$ & 2.17 & 3.04 & 3.72 & 4.28\\
$\widetilde{\mathbb{X}}^{(+)}\,[X(4630)]$ & $-1$ & 0.133 & 0.184 & 0.225 & 0.257\\
$\widetilde{\mathbb{X}}^{(-)}\,[X(4630)]$ & $+1$ & 0.120 & 0.166 & 0.203 & 0.232\\
$\widetilde{\mathbb{X}}^{(-)}\,[X(4630)]$ & $-1$ & 2.16 & 3.02 & 3.70 & 4.25\\
$\mathbb{X}^{\prime(-)}\,[Y(4626)]$ & $+1$ & 1.16 & 1.52 & 1.79 & 2.00\\
$\mathbb{X}^{\prime(-)}\,[Y(4626)]$ & $-1$ & 12.5 & 16.2 & 19.1 & 21.3\\
$\mathbb{X}^{\prime(+)}\,[Y(4626)]$ & $+1$ & 12.6 & 16.4 & 19.3 & 21.5\\
$\mathbb{X}^{\prime(+)}\,[Y(4626)]$ & $-1$ & 1.29 & 1.68 & 1.98 & 2.22\\
$\widetilde{\mathbb{X}}^{(-)}\,[Y(4626)]$ & $+1$ & 0.127 & 0.177 & 0.217 & 0.249\\
$\widetilde{\mathbb{X}}^{(-)}\,[Y(4626)]$ & $-1$ & 2.29 & 3.22 & 3.96 & 4.56\\
$\widetilde{\mathbb{X}}^{(+)}\,[Y(4626)]$ & $+1$ & 2.31 & 3.24 & 3.99 & 4.59\\
$\widetilde{\mathbb{X}}^{(+)}\,[Y(4626)]$ & $-1$ & 0.141 & 0.197 & 0.241 & 0.276\\
\bottomrule
\end{tabular}


\end{table*}

The relative sign between diagrams (a1) and (a2) has its largest effect in the scalar-vector assignments. For the experimentally anchored $\mathbb{X}^{(-)}[Y(4500)]$ at $\alpha=4$, the $s_{\mathrm a}=+1$ choice gives $\mathcal B_{\mathrm a}=4.12\times10^{-10}$, leaving $\mathcal B_{\mathrm b}=4.59\times10^{-7}$ to dominate the total. For $s_{\mathrm a}=-1$, the leading contributions add constructively and give $\mathcal B=1.36\times10^{-5}$. The secondary topology pair therefore matters chiefly near the cancellation of the larger individual diagrams.

For the axial reference assignments at $X(4630)$, $s_{\mathrm a}=+1$ gives the larger central branching fraction. The central ranges over $2\leq\alpha\leq5$ are $(3.23$--$5.50)\times10^{-5}$ for $\mathbb{X}^{\prime(+)}$ and $(2.17$--$4.28)\times10^{-6}$ for $\widetilde{\mathbb{X}}^{(+)}$. At the $Y(4626)$ label the larger branching fraction of each reference assignment instead has $s_{\mathrm a}=-1$, giving $(1.25$--$2.13)\times10^{-5}$ and $(2.29$--$4.56)\times10^{-6}$ for $\mathbb{X}^{\prime(-)}$ and $\widetilde{\mathbb{X}}^{(-)}$.

The five opposite-$C$ hypotheses give the following larger central branching fractions over the same cutoff range: $(8.59$--$17.2)\times10^{-6}$ with $s_{\mathrm a}=+1$ for $\mathbb{X}^{(+)}[Y(4500)]$; $(3.20$--$5.45)\times10^{-5}$ with $s_{\mathrm a}=-1$ for $\mathbb{X}^{\prime(-)}[X(4630)]$; $(2.16$--$4.25)\times10^{-6}$ with $s_{\mathrm a}=-1$ for $\widetilde{\mathbb{X}}^{(-)}[X(4630)]$; $(1.26$--$2.15)\times10^{-5}$ with $s_{\mathrm a}=+1$ for $\mathbb{X}^{\prime(+)}[Y(4626)]$; and $(2.31$--$4.59)\times10^{-6}$ with $s_{\mathrm a}=+1$ for $\widetilde{\mathbb{X}}^{(+)}[Y(4626)]$. The corresponding smaller branching fractions at selected cutoff values are given in Table~\ref{tab:BR4630}. In every reference--partner pair, the larger contribution from the leading topology pair moves to the opposite source sign, as required by Eq.~\eqref{eq:Csignexchange}.

For axial constituents, reversing $C$ also changes the contribution from the secondary topology class. It vanishes identically for all $C=-1$ axial assignments. For the $C=+1$ axial assignments at $\alpha=4$, it supplies about $1\%$ or less of the larger central branching fraction and about $10\%$ of the smaller one. This accounts for the small reference--partner differences beyond the interchange of the leading-source signs. The scalar-vector amplitudes in the secondary topology class do not obey the same antisymmetric relation. For $\mathbb{X}^{(+)}[Y(4500)]$ at $\alpha=4$, this pair supplies $11.0\%$ of the $s_{\mathrm a}=+1$ total and $99.97\%$ of the $s_{\mathrm a}=-1$ total; its two projected totals therefore are not obtained merely by interchanging the source signs of the reference assignment.

The input intervals induce asymmetric bands because the ERE couplings depend nonlinearly on the pole location and width, while the Weinberg couplings depend on the binding energy. Since the production amplitude is linear in a common $g_{\rm mol}$, the bands rescale each central curve without changing its cutoff dependence or the relative importance of the diagrams. An opposite-$C$ partner has the same fractional band as its reference assignment, not a new experimentally determined uncertainty. For the two Weinberg assignments, the bands reflect the quoted mass intervals alone; no additional width-induced uncertainty is implied.

The rates remain conditional on the factorized weak-source phase conventions, the common-velocity spin projectors, and the stated off-shell continuation. Heavy-quark symmetry of the strong vertices alone does not determine nonfactorizable weak-source phases. Further positive-parity sources and interactions within the excited doublets are not included. For the partner calculations, possible splittings of masses, widths and molecular residues are likewise outside the present minimal assumption. Their effects, as well as variations of the propagator and weak off-shell prescriptions, are outside the coupling-only envelopes.

\section{Summary}\label{sec:summary}
We have studied $B_s^0$ decays into $\eta$ and spin-one molecular candidates composed of $D_{s0}^*(2317)\bar D_s^*$, $D_{s1}(2460)\bar D_s^*$ or $D_{s1}(2536)\bar D_s^*$. The production amplitudes combine factorized weak vertices, heavy-hadron chiral transitions and threshold-dependent constituent-coupling estimates. Above threshold we used an exploratory ERE compositeness input together with two-channel width saturation; below the $D_{s1}(2536)\bar D_s^*$ threshold we used Weinberg's weak-binding relation with $X_{\rm mol}=1$. The corresponding uncertainty rules are, respectively, nonnegative mass--width feasible-domain envelopes and mass-only binding-energy variation. In addition to five experimentally anchored assignments, we calculated five opposite-$C$ partner hypotheses with the same benchmark masses, widths and molecular couplings. These partners and their rates are conditional on this degeneracy prescription, not additional experimental identifications or independent coupling fits.

Among the three classes of triangle mechanisms, the dominant contributions come from the leading pair, diagrams (a1) and (a2), whose relative sign strongly affects the rate. The secondary topology pair, diagrams (b1) and (b2), becomes important mainly when the leading contributions cancel, whereas the contractions in the third class, diagrams (c1) and (c2), vanish for the retained tensor structures. Reversing $C$ interchanges the two relative-sign choices of the leading source pair. For axial constituents it also changes the secondary topology pair from exact cancellation at $C=-1$ to a finite, generally subleading contribution at $C=+1$. With no interference between the complete leading and secondary amplitudes, the larger central branching fractions span several $10^{-6}$ to several $10^{-5}$ over $2\leq\alpha\leq5$.

Beyond these numerical estimates, this study connects weak $B_s^0$ production with the internal structure of hidden-charm, hidden-strangeness candidates. The charge-conjugation projections show that configurations built from the same meson pair can have distinct production patterns even when their benchmark masses, widths and molecular couplings are held fixed. The cutoff curves and inherited coupling bands therefore provide reference scales for future high-statistics searches involving an $\eta$ and the corresponding charm-strange decay products. Together with studies in other production environments, such measurements or upper limits could help test molecular assignments, constrain the relative weak-source phases, and clarify the role of opposite-$C$ partners in the $c\bar c s\bar s$ spectrum.

\begin{acknowledgments}
\label{sec:acknowledgements}
This work is supported by the National Natural Science Foundation of China under Grants No.~12235018 and No.~12675106.
\end{acknowledgments}

\appendix
\onecolumngrid
\section{Weak-current and strong-transition tensors}\label{app:vertices}\label{app:strongmol}
\subsection{Transition currents}
For a transition meson with momentum $k_t$ and an emitted meson with momentum $k_e$, put
\begin{equation}
k_e=p_0-k_t,\qquad t=m_e^2.
\label{eq:weakmomenta}
\end{equation}
The weak current is $J_\mu=\bar b\gamma_\mu(1-\gamma_5)c$. For a spin-one transition meson, define $\langle D_s^{*-}|J_\mu|B_s^0\rangle=\epsilon_t^{*\nu}J_{D_s^*;\mu\nu}$, and analogously for $D_{s1}^{(\prime)}$. Both tensor indices are lower: the first is the current index and the second contracts the transition polarization. In the phase convention of Ref.~\cite{Cheng:2003sm},
\begin{align}
J_{D_s;\mu}&=F_1^{B_sD_s}(t)(p_0+k_t)_\mu+
\frac{m_{B_s}^2-m_{D_s}^2}{t}[F_0^{B_sD_s}(t)-F_1^{B_sD_s}(t)]k_{e\mu},
\label{eq:PPff}\\
J_{D_{s0}^*;\mu}&=i\left\{F_1^{B_sD_{s0}^*}(t)(p_0+k_t)_\mu+
\frac{m_{B_s}^2-m_{D_{s0}^*}^2}{t}[F_0^{B_sD_{s0}^*}(t)-F_1^{B_sD_{s0}^*}(t)]k_{e\mu}\right\},
\label{eq:PSff}\\
J_{D_s^*;\mu\nu}&=-\frac{A^{B_sD_s^*}(t)}{m_{B_s}+m_{D_s^*}}
\varepsilon_{\mu\nu\rho\sigma}(p_0+k_t)^\rho k_e^\sigma
-i\Biggl[(m_{B_s}+m_{D_s^*})A_1(t)g_{\mu\nu}\nonumber\\
&\qquad-\frac{A_2(t)}{m_{B_s}+m_{D_s^*}}(p_0+k_t)_\mu(p_0+k_t)_\nu
-\frac{2m_{D_s^*}}{t}[A_3(t)-A_0(t)]k_{e\mu}(p_0+k_t)_\nu\Biggr],
\label{eq:PVff}\\
J_{D_{s1}^{(\prime)};\mu\nu}&=+\frac{A^{B_sD_{s1}^{(\prime)}}(t)}{m_{B_s}-m_{D_{s1}^{(\prime)}}}
\varepsilon_{\mu\nu\rho\sigma}(p_0+k_t)^\rho k_e^\sigma
-i\Biggl[(m_{B_s}-m_{D_{s1}^{(\prime)}})V_1(t)g_{\mu\nu}\nonumber\\
&\qquad-\frac{V_2(t)}{m_{B_s}-m_{D_{s1}^{(\prime)}}}(p_0+k_t)_\mu(p_0+k_t)_\nu
-\frac{2m_{D_{s1}^{(\prime)}}}{t}[V_3(t)-V_0(t)]k_{e\mu}(p_0+k_t)_\nu\Biggr].
\label{eq:PAff}
\end{align}
The form factors in the last equation refer to the specified $D_{s1}'$ or $D_{s1}$ transition, without mixing their inputs. The functions $A_i$ describe the $D_s^*$ transition, and $V_i$ the selected axial transition. The scalar coefficients are evaluated at $t=m_e^2$, with the loop-dependent four-vectors retained.

The auxiliary functions are
\begin{align}
A_3(t)&=\frac{m_{B_s}+m_{D_s^*}}{2m_{D_s^*}}A_1(t)-\frac{m_{B_s}-m_{D_s^*}}{2m_{D_s^*}}A_2(t),
\label{eq:A3}\\
V_3(t)&=\frac{m_{B_s}-m_{D_{s1}^{(\prime)}}}{2m_{D_{s1}^{(\prime)}}}V_1(t)-\frac{m_{B_s}+m_{D_{s1}^{(\prime)}}}{2m_{D_{s1}^{(\prime)}}}V_2(t).
\label{eq:V3}
\end{align}
Exact analytic parameterizations satisfy $A_3(0)=A_0(0)$ and $V_3(0)=V_0(0)$, ensuring regularity at $t=0$. As specified in Sec.~\ref{sec:weakvertices}, the rounded fits in Table~\ref{tab:formfactors} are used only as phenomenological inputs at $t=m_e^2$, not as complete $t$-dependent currents.

\subsection{Strong transitions}
Put $c_\eta=-\sqrt{2/3}$ and write all vertices in terms of the $\eta$ four-momentum $p_1$, the common velocity $v$ and the metric $g_{\mu\nu}$. With the charged-field phases of Eq.~\eqref{eq:Cphases}, the real reduced tensors are
\begin{align}
E_{D_sD_s^*,\mu}&=c_\eta\frac{2g_{\rm HH}\sqrt{m_{D_s}m_{D_s^*}}}{f_\pi}p_{1\mu},\nonumber\\
E_{D_s^*D_s^*,\rho\beta}&=c_\eta\frac{2g_{\rm HH} m_{D_s^*}}{f_\pi}
\varepsilon_{\rho\sigma\alpha\beta}p_1^\sigma v^\alpha,
\label{eq:HHchi}\\
E_{D_{s0}^*D_s}&=c_\eta\frac{2h\sqrt{m_{D_{s0}^*}m_{D_s}}}{f_\pi}(v\cdot p_1),\nonumber\\
E_{D_{s1}'D_s^*,\mu\nu}&=-c_\eta\frac{2h\sqrt{m_{D_{s1}'}m_{D_s^*}}}{f_\pi}
\nonumber\\ &\quad\times(v\cdot p_1)(g_{\mu\nu}-v_\mu v_\nu),\nonumber\\
E_{D_{s1}D_s^*,\mu\nu}&=-c_\eta\frac{4h'\sqrt{m_{D_{s1}}m_{D_s^*}}}{\sqrt6 f_\pi\Lambda_\chi}
\nonumber\\ &\quad\times\Bigl[3p_{1\mu}p_{1\nu}
-3(v\cdot p_1)(p_{1\mu}v_\nu+v_\mu p_{1\nu})\nonumber\\
&\qquad+g_{\mu\nu}\bigl[(v\cdot p_1)^2-p_1^2\bigr]\nonumber\\
&\qquad+v_\mu v_\nu\bigl[2(v\cdot p_1)^2+p_1^2\bigr]\Bigr].
\label{eq:STlag}
\end{align}
The heavy external-state factors $\sqrt{m_im_j}$ convert the heavy-field normalization to the relativistic normalization used for the weak and molecular vertices.

\section{Triangle loop amplitudes}\label{app:loops}
The common triangle functional is defined in Eq.~\eqref{eq:loopgeneric}. Table~\ref{tab:routing} gives the line assignments and weak-current ordering used below.

\begin{table}[ht]
\caption{Upper, lower and exchanged particles for each panel, with the electric charge of every internal charmed-strange meson shown explicitly. The weak-tensor arguments are the transition and emitted momenta, respectively. For $\widetilde{\mathbb{X}}$ replace $D_{s1}^{\prime\pm}$ by $D_{s1}^{\pm}$ in the lower block and in the corresponding weak tensors.}
\label{tab:routing}\centering\small
\begin{tabular}{clllll}\toprule
Molecule & Panel & Upper $r$ & Lower $q$ & Exchange $\ell$ & Weak tensor\\\midrule
$\mathbb{X}$ & $\mathrm{a1}$ & $D_s^-$ & $D_{s0}^{*+}$ & $D_s^{*-}$ & $\mathcal W_{D_s,D_{s0}^*}(r,q)$\\
& $\mathrm{a2}$ & $D_s^+$ & $D_{s0}^{*-}$ & $D_s^{*+}$ & $\mathcal W_{D_{s0}^*,D_s}(q,r)$\\
& $\mathrm{b1}$ & $D_s^+$ & $D_s^{*-}$ & $D_{s0}^{*+}$ & $\mathcal W_{D_s^*,D_s}(q,r)$\\
& $\mathrm{b2}$ & $D_s^-$ & $D_s^{*+}$ & $D_{s0}^{*-}$ & $\mathcal W_{D_s,D_s^*}(r,q)$\\
& $\mathrm{c1}$ & $D_s^{*-}$ & $D_{s0}^{*+}$ & $D_s^{*-}$ & $\mathcal W_{D_s^*,D_{s0}^*}(r,q)$\\
& $\mathrm{c2}$ & $D_s^{*+}$ & $D_{s0}^{*-}$ & $D_s^{*+}$ & $\mathcal W_{D_{s0}^*,D_s^*}(q,r)$\\\midrule
$\mathbb{X}'$ & $\mathrm{a1}$ & $D_s^{*-}$ & $D_{s1}^{\prime+}$ & $D_s^{*-}$ & $\mathcal W_{D_s^*,D_{s1}'}(r,q)$\\
& $\mathrm{a2}$ & $D_s^{*+}$ & $D_{s1}^{\prime-}$ & $D_s^{*+}$ & $\mathcal W_{D_{s1}',D_s^*}(q,r)$\\
& $\mathrm{b1}$ & $D_s^{*+}$ & $D_s^{*-}$ & $D_{s1}^{\prime+}$ & $\mathcal W_{D_s^*,D_s^*}(q,r)$\\
& $\mathrm{b2}$ & $D_s^{*-}$ & $D_s^{*+}$ & $D_{s1}^{\prime-}$ & $\mathcal W_{D_s^*,D_s^*}(r,q)$\\
& $\mathrm{c1}$ & $D_s^-$ & $D_{s1}^{\prime+}$ & $D_s^{*-}$ & $\mathcal W_{D_s,D_{s1}'}(r,q)$\\
& $\mathrm{c2}$ & $D_s^+$ & $D_{s1}^{\prime-}$ & $D_s^{*+}$ & $\mathcal W_{D_{s1}',D_s}(q,r)$\\\bottomrule
\end{tabular}\end{table}
The three particle names in its subscript specify the upper, lower and exchanged masses, in that order. For example, $\mathcal I_{D_s,D_{s0}^*,D_s^*}$ uses $D_s$ on the upper line, $D_{s0}^*$ on the lower line and $D_s^*$ on the exchange. The table and all contractions below use the panel numbering of Figs.~\ref{fig:Ds0loops} and~\ref{fig:Ds1loops}. The strong tensors $E$ and molecular tensors $U$ are given in Eqs.~\eqref{eq:HHchi}, \eqref{eq:STlag} and \eqref{eq:Umol}. All repeated Lorentz indices are contracted covariantly. The amplitudes below are defined before charge projection; the molecular charge coefficients are inserted once through Eq.~\eqref{eq:phaseamp}.

\paragraph{Production of $\mathbb{X}$.}
The six charged contractions are
\begin{align}
\mathcal A_{\mathrm{a1}}&=\mathcal I_{D_s,D_{s0}^*,D_s^*}\!\left[
\mathcal W_{D_s,D_{s0}^*}(r,q)E_{D_sD_s^*,\mu}\Pi^{\mu\nu}(v)U_{\mathbb{X},\nu}\right],
\label{eq:loopDs0a}\\
\mathcal A_{\mathrm{a2}}&=-\mathcal I_{D_s,D_{s0}^*,D_s^*}\!\left[
\mathcal W_{D_{s0}^*,D_s}(q,r)E_{D_sD_s^*,\mu}\Pi^{\mu\nu}(v)U_{\mathbb{X},\nu}\right],
\label{eq:loopDs0b}\\
\mathcal A_{\mathrm{b1}}&=\mathcal I_{D_s,D_s^*,D_{s0}^*}\!\left[
\mathcal W_{D_s^*,D_s;\mu}(q,r)\Pi^{\mu\nu}(v)U_{\mathbb{X},\nu}E_{D_{s0}^*D_s}\right],
\label{eq:loopDs0cp}\\
\mathcal A_{\mathrm{b2}}&=\mathcal I_{D_s,D_s^*,D_{s0}^*}\!\left[
\mathcal W_{D_s,D_s^*;\mu}(r,q)\Pi^{\mu\nu}(v)U_{\mathbb{X},\nu}E_{D_{s0}^*D_s}\right],
\label{eq:loopDs0cm}\\
\mathcal A_{\mathrm{c1}}&=\mathcal I_{D_s^*,D_{s0}^*,D_s^*}\!\left[
\mathcal W_{D_s^*,D_{s0}^*;\mu}(r,q)\Pi^{\mu\alpha}(v)
E_{D_s^*D_s^*,\beta\alpha}\Pi^{\beta\nu}(v)U_{\mathbb{X},\nu}\right],
\label{eq:loopDs0dp}\\
\mathcal A_{\mathrm{c2}}&=\mathcal I_{D_s^*,D_{s0}^*,D_s^*}\!\left[
\mathcal W_{D_{s0}^*,D_s^*;\mu}(q,r)\Pi^{\mu\alpha}(v)
E_{D_s^*D_s^*,\beta\alpha}\Pi^{\beta\nu}(v)U_{\mathbb{X},\nu}\right].
\label{eq:loopDs0dm}\end{align}
The minus in diagram (a2) is the fixed charged $D_sD_s^*\eta$ vertex sign, separate from the weak-source phase. In pair b, the exchanged $D_{s0}^*$ has no spin projector. The weak tensor for diagram (b1) has a $D_s^*$ transition and $D_s$ emission; diagram (b2) has a $D_s$ transition and $D_s^*$ emission.
Thus (b1) feeds $D_{s0}^{*+}D_s^{*-}$ and (b2) feeds $D_{s0}^{*-}D_s^{*+}$, in the same charge order as the a and c pairs.

\paragraph{Production of $\mathbb{X}'$ and $\widetilde{\mathbb{X}}$.}
For $\mathbb{X}'$ the contractions are
\begin{align}
\mathcal A_{\mathrm{a1}}&=\mathcal I_{D_s^*,D_{s1}',D_s^*}\!\left[
\mathcal W_{D_s^*,D_{s1}';\mu\nu}(r,q)\Pi^{\nu\beta}(v)
E_{D_s^*D_s^*,\rho\beta}\Pi^{\mu\gamma}(v)\Pi^{\rho\delta}(v)U_{\mathbb{X}',\gamma\delta}\right],
\label{eq:loopDs1a}\\
\mathcal A_{\mathrm{a2}}&=\mathcal I_{D_s^*,D_{s1}',D_s^*}\!\left[
\mathcal W_{D_{s1}',D_s^*;\mu\nu}(q,r)\Pi^{\mu\beta}(v)
E_{D_s^*D_s^*,\rho\beta}\Pi^{\nu\gamma}(v)\Pi^{\rho\delta}(v)U_{\mathbb{X}',\gamma\delta}\right],
\label{eq:loopDs1b}\\
\mathcal A_{\mathrm{b1}}&=-\mathcal I_{D_s^*,D_s^*,D_{s1}'}\!\left[
\mathcal W_{D_s^*,D_s^*;\mu\nu}(q,r)\Pi^{\mu\alpha}(v)
E_{D_{s1}'D_s^*,\beta\alpha}\Pi^{\beta\gamma}(v)\Pi^{\nu\delta}(v)U_{\mathbb{X}',\gamma\delta}\right],
\label{eq:loopDs1cp}\\
\mathcal A_{\mathrm{b2}}&=\mathcal I_{D_s^*,D_s^*,D_{s1}'}\!\left[
\mathcal W_{D_s^*,D_s^*;\mu\nu}(r,q)\Pi^{\nu\alpha}(v)
E_{D_{s1}'D_s^*,\beta\alpha}\Pi^{\beta\gamma}(v)\Pi^{\mu\delta}(v)U_{\mathbb{X}',\gamma\delta}\right],
\label{eq:loopDs1cm}\\
\mathcal A_{\mathrm{c1}}&=\mathcal I_{D_s,D_{s1}',D_s^*}\!\left[
\mathcal W_{D_s,D_{s1}';\mu}(r,q)\Pi^{\mu\gamma}(v)
E_{D_sD_s^*,\nu}\Pi^{\nu\delta}(v)U_{\mathbb{X}',\gamma\delta}\right],
\label{eq:loopDs1dp}\\
\mathcal A_{\mathrm{c2}}&=-\mathcal I_{D_s,D_{s1}',D_s^*}\!\left[
\mathcal W_{D_{s1}',D_s;\mu}(q,r)\Pi^{\mu\gamma}(v)
E_{D_sD_s^*,\nu}\Pi^{\nu\delta}(v)U_{\mathbb{X}',\gamma\delta}\right].
\label{eq:loopDs1dm}\end{align}
The $\widetilde{\mathbb{X}}$ amplitudes follow by replacing every $D_{s1}'$ by $D_{s1}$, including its mass and weak inputs, $E_{D_{s1}'D_s^*}$ by $E_{D_{s1}D_s^*}$, and $U_{\mathbb{X}'}$ by $U_{\widetilde{\mathbb{X}}}$. This replacement changes the $S$-wave strong emission to the stated $D$-wave emission in pair b.

In pair b, the first index of $E_{D_{s1}'D_s^*}$ is the exchanged axial index and the second the upper vector index; these symmetric tensors retain this label order. The minus signs in diagrams (b1) and (c2) are the fixed charged $D_{s1}'D_s^*\eta$ and $D_sD_s^*\eta$ signs, respectively. The two b contractions share one $D_s^{*+}D_s^{*-}$ weak source and have no independent weak phases, including after the replacements for $\widetilde{\mathbb{X}}$.

\twocolumngrid
\bibliography{references}

\end{document}